\documentclass[12pt]{article}
\usepackage[USenglish,activeacute]{babel}
\usepackage{amsfonts,amssymb,amsmath}
\usepackage{color}

\numberwithin{equation}{section}

\begin{document}

\title{\textbf{Mass in Lovelock Unique Vacuum }\\
\textbf{gravity theories}}
\author{Gabriel Arenas-Henriquez $^{1,}$\thanks{gabriel.arenas.henriquez@gmail.com} ,
Robert B. Mann $^{2,3,}$\thanks{rbmann@uwaterloo.ca} , \\
Olivera Miskovic $^{4,}$\thanks{olivera.miskovic@pucv.cl} \
and Rodrigo Olea $^{1,}$\thanks{rodrigo.olea@unab.cl}\bigskip \\
{\small \emph{$^1$Departamento de Ciencias F\'{\i}sicas, Universidad Andres Bello,}}\\
{\small \emph{Sazi\'{e} 2212, Piso 7, Santiago, Chile}}\\
{\small \emph{$^2$Department of Physics and Astronomy, University of
Waterloo,}}\\
{\small \emph{Waterloo, Ontario, N2L 3G1, Canada}}\\
{\small \emph{$^3$Perimeter Institute, 31 Caroline Street North,}}\\
{\small \emph{Waterloo, ON, N2L 2Y5, Canada}}\\
{\small \emph{$^4$Instituto de F\'\i sica, Pontificia Universidad Cat\'olica
de Valpara\'\i so,}}\\
{\small \emph{Casilla 4059, Valpara\'{\i}so, Chile}} }
\maketitle

\abstract{We derive an expression for conserved charges in
 Lovelock AdS gravity for solutions having $k$-fold degenerate vacua, making manifest a link between
the degeneracy of a given vacuum  and the nonlinearity of the energy formula.  We show  for
a black hole solution to the field equations on a branch of multiplicity $k$ that its mass comes from an expression that contains the product of $k$ Weyl tensors.
We prove that all divergent contributions of the type (Weyl)$^q$, with $1\le q<k$, are suppressed. Our conserved charge definition is a natural generalization of the Conformal Mass by Ashtekar, Magnon and Das to the cases when $k>1$.  Our results provide insight on the holographic properties of degenerate Lovelock theories.
}

\section{Introduction}

In the ongoing effort to construct an ultraviolet complete theory of quantum gravity, higher curvature
theories of gravity emerge as corrections to the usual Einstein- Hilbert term.  Such theories have now
 come to play an important role in  cosmology, black hole physics, supergravity, string theory and holography.   They provide a framework for understanding which features of gravitational theory
are generic and which are special.    Amongst the most important of these theories is  Lovelock gravity \cite{Lovelock:1971yv}, which  is the natural generalization  of General Relativity to $D$ dimensions in two important respects: its field equations are of second-order and it is  free of ghosts when expanded on a background of constant curvature  \cite{Zwiebach:1985uq}.

One of the most important questions to address in any gravitational theory is the definition of conserved charges.  This is a subtle issue in gravitational physics because of the equivalence principle, which makes localization of gravitational energy (and momentum and angular momentum) fraught with ambiguities \cite{MTW}.  Approaches for addressing this problem date back six decades
\cite{Arnowitt:1960es,Arnowitt:1962hi} and more, and include a mixture of global \cite{Ashtekar:1984zz,Balasubramanian:1999re,Ashtekar:1999jx,Das:2000cu,Mann:2005yr} and quasilocal  \cite{Brown:1992bq,Brown:1994gs,Creighton:1995au,McGrath:2012db,Epp:2013hua} methods.     Generalizations
to Lovelock gravity have been carried out \cite{Olea:2006vd}, and a
universal form of the boundary term yielding a background-independent definition of conserved quantities for any Lovelock gravity theory with anti-de Sitter (AdS) asymptotics has been constructed \cite{Kofinas:2008ub}.

The purpose of this paper is to extend our understanding of conserved charges in Lovelock gravity to degenerate solutions of the field equations.  The simplest form of such solutions is that of a $k$-fold degenerate AdS vacuum. In general the field equations of  $N$-th order Lovelock  gravity  admit
as many as $N$ distinct vacua of constant curvature, each with their own effective cosmological constant.  However these vacua need not be distinct -- depending on the values of the Lovelock coupling
constants, as many as $k$ cosmological constants can be equal to each other, where  $k \leq N$; when
this takes place we say that the vacuum is $k$-fold degenerate.  This concept extends to solutions, such as black holes, that do not have constant curvature.

Understanding what the conserved charges are for such solutions is rather subtle.  The asymptotic falloff conditions differ from those cases in which the solution has no degeneracy, and standard approaches
\cite{Balasubramanian:1999re,Ashtekar:1999jx,Das:2000cu} do not work.

In this paper we address this problem. We obtain an expression for conserved charges in
 Lovelock AdS gravity for  $k$-fold degenerate solutions. In particular we demonstrate that the mass of
 a black hole solution to the field equations on a branch of multiplicity $k$ comes from an expression containing the product of $k$ Weyl tensors.  There is thus a  link between
the degeneracy of a given vacuum  and the nonlinearity of the energy formula.   We find that
potentially divergent contributions of the type (Weyl)$^q$, with $1\le q<k$, are in fact suppressed.
Noting the obstruction to linearization of such solutions about a constant curvature background
\cite{Fan:2016zfs,Camanho:2013pda}, our results can provide
useful insight on the holographic properties of degenerate Lovelock gravity. This
may be regarded as a natural generalization of Conformal Mass to these
theories \cite{Ashtekar:1999jx,Das:2000cu}.

The outline of our paper is as follows. We begin in section \ref{sec2} with a review of Lovelock gravity,
highlighting the notion of degenerate solutions.  In section \ref{sec3} we review the construction
of conserved charges in the Kounterterm formalism \cite{Olea:2006vd}, and in section \ref{sec4} we
derive a general formula for conserved charges for degenerate solutions (technical details are shown in Appendix A). Although even and odd dimensions need to be treated distinctly, we find that a single expression is valid in any dimension,
and discuss its applications to obtaining the mass of a black hole solution.  We summarize our
work in section \ref{sec5}.

\section{Lovelock AdS gravity}
\label{sec2}

\label{Lovelock} Lovelock theory of gravity is the most general theory whose
dynamics is described by a second-order equation of motion. The action on a $%
D$-dimensional manifold $\mathcal{M}$ endowed by the metric $g_{\mu \nu}$ is
a polynomial in the Riemann curvature,
\begin{equation}
I = \frac { 1 }{ 16\pi G } \int\limits_{\mathcal{M}}{\ d^{ D }x\sqrt { -g }
\sum _{ p=0 }^{ N } \alpha _{ p }\mathcal{L}_p(R)}\,,  \label{action}
\end{equation}
where $\alpha_{p}$ are the coupling constants, $N=[(D-1)/2]$ and $\mathcal{L}%
_{p}$ is the dimensionally continuation of the $p$-th Euler density,
\begin{equation}
\mathcal{L}_p =\frac{1}{2^p}\,\delta _{\nu_1\cdots \nu _{2p}}^{\mu_1\cdots
\mu _{2p}}\,R_{\mu _{1}\mu _{2}}^{\nu _{1}\nu _{2}}\cdots R_{\mu_{2p-1}\mu
_{2p}}^{\nu _{2p-1}\nu _{2p}}\,.
\end{equation}
Here, $\delta _{\nu _{1}\cdots \nu _{m}}^{\mu _{1}\cdots \mu _{m}}=\det
[\delta _{\nu _{1}}^{\mu _{1}}\cdots \delta _{\nu _{m}}^{\mu _{m}}] $
denotes a completely antisymmetric Kronecker delta of rank $m$. We can think
of this series as a modification to the Einstein-Hilbert action with
negative cosmological constant $\Lambda=-(D-1)(D-2)/2\ell^{2}$ with $\ell$
as the bare AdS radius. The Einstein-Hilbert term corresponds to first two
terms in the polynomial (\ref{action}), with $\alpha_{0}=-2\Lambda$ and $%
\alpha_{1}=1$. Then the variation of the action leads to the following
equations of motion,
\begin{eqnarray}
\mathcal{E}_\mu^\nu=R_{ \mu }^{ \nu }-\frac { 1 }{ 2 } R{\ \delta }_{ \mu
}^{ \nu }+\Lambda \delta _{ \mu }^{ \nu }-H_{ \mu }^{ \nu }=0\,,  \label{eom}
\end{eqnarray}
where $H^{\nu}_{\mu}$ is the generalized Lanczos tensor that contains all
the higher-curvature contributions to the Einstein tensor in $D>4$,
\begin{equation}
H_\mu^\nu=\sum _{ p=2 }^{ N } \frac { \alpha _{ p } }{ 2^{ p+1 } }\, \delta
_{ \mu \mu _{ 1 }\cdots \mu _{ 2p } }^{ \nu \nu _{ 1 }\cdots \nu _{ 2p } }\,
R_{ \nu _{ 1 }\nu _{ 2 } }^{ \mu _{ 1 }\mu _{ 2 } }\cdots R_{ \nu _{ 2p-1
}\nu _{ 2p } }^{ \mu _{ 2n-1 }\mu _{ 2p }}\,.  \label{LL}
\end{equation}

\subsection{Vacua of the theory}

An alternative way to write the field equations is
\begin{align}
\mathcal{E}_\mu^\nu &= \sum _{ p=0}^{ N } \frac { \alpha _{ p } }{ 2^{ p+1 } } \delta _{ \mu \mu _{ 1 }\cdots \mu _{ 2p } }^{
\nu \nu _{ 1 }\cdots \nu _{ 2p } }\, \left(R_{\nu _{ 1 }\nu _{ 2 } }^{ \mu
_{ 1 }\mu _{ 2 } }+\frac{1}{\ell_{\mathrm{eff}}^{(1) 2 }} \,\delta_{\nu _{ 1
}\nu _{ 2 } }^{ \mu _{ 1 }\mu _{ 2 } } \right)\cdots \left(R_{ \nu _{ 2p-1
}\nu _{ 2p } }^{ \mu _{ 2n-1 }\mu _{ 2p }}+\frac{1}{\ell_{\mathrm{eff}}^{(N)
2}}\,\delta_{ \nu _{ 2p-1 }\nu _{ 2p } }^{ \mu _{ 2n-1 }\mu _{ 2p }}
\right) \nonumber\\
&=0\,,  \label{product of AdS}
\end{align}
where
$\ell_{\mathrm{eff}}^{(i)}$, with $i=1,\cdots,N$, are effective AdS
radii. They differ from the bare AdS radius $\ell$ due to the contribution
of higher-order terms through the couplings $\{\alpha_2,\ldots , \alpha_N\}$.
Eq.(\ref{product of AdS}) explicitly shows that maximally-symmetric
spacetimes are particular  vacuum solutions of the theory, whose Riemann
curvature is
\begin{equation}
R^{\mu \nu}_{\alpha \beta}=-\frac{1}{\ell_{\mathrm{eff}}^{2}}\,\delta^{\mu
\nu}_{\alpha \beta}   \label{maxR}
\end{equation}
which we can also write as $R+\frac{1}{\ell _{\mathrm{eff}}^{2}}
\delta^{[2]}  = 0$, suppressing indices using the square-bracket notation.
We can find the form of $\ell_{\mathrm{eff}}$ by inserting the above
expression into $\mathcal{E}_\mu^\nu=0$ and obtain the characteristic
polynomial of order $N$ in the variable $\ell _{\mathrm{eff}}^{-2}$,
\begin{equation}
\Delta (\ell _{\mathrm{eff}}^{-2}) \equiv \sum_{p=0}^{N}{\frac{%
(D-3)!\,(-1)^{p-1}{\alpha }_{p}}{\left( D-2p-1\right) !}}\left( \frac{1}{%
\ell _{\mathrm{eff}}^{2}}\right) ^{p}=0\,.  \label{Delta}
\end{equation}
The roots of $\Delta$ depend on the parameter set $\{\alpha_p\}$; the simplest
case is $N=1$  with $\Lambda = -(D-1)(D-2)/2/\ell _{\mathrm{eff}}^{2}$, for which
$\ell _{\mathrm{eff}} = \ell$.   Writing $\lambda=\ell _{\mathrm{eff}}^{-2}$, if there
are $N$ real roots $\lambda_i$ ($i=1\ldots N$) then the theory has $%
N$ maximally symmetric vacua, which can be AdS ($\lambda_i >0$), dS ($\lambda_i <0$%
) or flat ($\lambda_i =0$). If a root $\lambda_i$ is complex then a maximally
symmetric space for this value of $\ell _{\mathrm{eff}}$
is not a particular solution of the theory.

We are interested in a spacetime that is asymptotically AdS, defined as a
branch of the theory by the corresponding AdS radius $\ell_{\mathrm{eff}}$.
The root itself can be simple, or $k$-fold degenerate, describing a vacuum of
multiplicity $k$.

The existence of a vacuum with   multiplicity $k>1$ is a defining feature
of degenerate theories. A particular root is simple if its characteristic
polynomial satisfies
\begin{equation}
\Delta^{(1)}(\ell _{\mathrm{eff}}^{-2})=\frac{d\Delta}{d(\ell_{\mathrm{eff}}^{-2})}=\sum_{p=1}^{N}
\frac{(D-3)!\,(-1)^{p-1}p\alpha_{p}}{(D-2p-1)!} \left(\frac{1}{\ell_{\mathrm{eff}%
}^{2}}\right)^{p-1}\neq 0\,.  \label{DeltaPrime}
\end{equation}
The relation $\Delta^{(1)}=0$ defines a \textit{critical} point in the
parameter space, which acts as an obstruction to the linearization of the
theory (see, e.g, ref.\cite{Fan:2016zfs} in the context of
Einstein-Gauss-Bonnet gravity and ref.\cite{Camanho:2013pda} in Lovelock
gravity). This fact prevents the obtention of an energy formula as a linearized charge \cite{Deser:2002jk,Petrov:2019roe}.
 The nature of this criticality will depend on how degenerate the
particular branch is.

We will focus on a theory with a vacuum that is $k$-fold degenerate. It is defined
by the fact that all derivatives of $\Delta$ vanish up to order $k$,
\begin{eqnarray}
\Delta^{(q)}(\ell_{\mathrm{eff}}^{-2})&=&{\frac{1}{q!}\frac{d^q\Delta}{d(\ell_{\mathrm{eff}}^{-2})^q}}=0\,,\qquad \mathrm{when} \quad 1 \leq q \leq
k-1\,,  \notag \\
\Delta^{(k)}(\ell _{\mathrm{eff}}^{-2})&=& \sum_{p=k}^{N}\binom{p}{k}\frac{%
(D-3)!\,(-1)^{p-1}\alpha_{p}\,(\ell _{\mathrm{eff}}^{-2})^{p-k}}{(D-2p-1)!}%
\neq 0\,.  \label{DeltaK}
\end{eqnarray}
{The above condition means that the equations of motion (\ref{product of AdS}%
) are factorizable by $\left( R+\frac{1}{\ell _{\mathrm{eff}}^{2}}%
\delta^{[2]}\right) ^{k}$.}

Examples of degenerate theories with maximal multiplicity are, in even
dimensions $D=2n$, Born-Infeld AdS gravity,
\begin{equation}
\alpha_p^{\mathrm{BI}}=\binom{n-1}{p}\frac{2^{n-2}(2n-2p-1)!}{\ell^{2(n-p-1)}%
} \,,\qquad 0\leq p\leq n-1 \,,
\end{equation}
and, in odd dimensions $D=2n+1$, Chern-Simons AdS gravity,
\begin{equation}
\alpha_p^{\mathrm{CS}}=\binom{n}{p}\frac{2^{n-1}(2n-2p)!}{\ell^{2(n-p)}}
\,,\qquad 0\leq p\leq n \, \,.
\end{equation}

A special feature of Chern-Simons AdS gravity is that it has enhanced local symmetry from Lorentz to AdS, so its dynamics and a number of degrees of freedom are different compared to any other Lovelock gravity.
In that sense, Chern-Simons AdS gravity is a truly gauge theory. Similar features hold for Chern-Simons de Sitter and Poincar\'e gravities.

A generalization of the above theories, which considers a unique solution for  $\ell_{\mathrm{eff}}$ with an intermediate degeneracy $k$, is the so-called Lovelock Unique Vacuum (LUV) AdS theory in $D$ dimensions. Here, $\Delta ^{(k)}$ is the only nonvanishing coefficient meaning that, apart from
(\ref{DeltaK}), there is also the additional requirement $\Delta ^{(q)}(\ell
_{\mathrm{eff}}^{-2})=0$ when $k<q\leq N$. In this case, the coupling
constants have the form
\begin{equation}
\alpha_p^{\mathrm{LUV}}=\binom{k}{p}\frac{2^{k-1}(D-2p-1)!}{\ell^{2(k-p)}}
  \qquad 0\leq p\leq k ,  \qquad  \alpha_{p > k}^{\mathrm{LUV}} = 0 \,. \label{alpha_LUV}
\end{equation}
Einstein-Hilbert, Born-Infeld and Chern-Simons AdS gravities are particular
cases of LUV AdS theories.

Our present goal is to find a mass formula for all those gravity theories
where the degeneracy is higher than one in odd and even spacetime
dimensions. To this end, we have to analyze the asymptotic behavior of the
corresponding solutions. In the next section, we obtain the falloff of the
relevant quantities for static geometries with $1 \leq k \leq N$.

\subsection{Asymptotic behavior of Lovelock AdS solutions}

\label{asymptotic}

A topological, static black hole solution in the local coordinates $x^\mu
=(t,r,y^m)$ ($m=2,\cdots, D-1$) is described by the line element
\begin{equation}
{\ d }s^{ 2 }=-f(r){\ dt }^{ 2 }+\frac { 1 }{ f(r) } {\ dr }^{ 2 }+{\ r }%
^{2} \gamma_{mn}(y)dy^mdy^n\,,  \label{ansatz}
\end{equation}
where $\gamma_{nm}$ is the metric of the transverse section of constant
curvature $\kappa=+1,0,-1$. The metric function $f(r)$ is given by the
algebraic master equation obtained as the first integral of the equations of
motion \cite{Boulware-Deser,CaiEGB,CaiLovelock},
\begin{equation}
\sum _{p=0}^{N}\,\frac{\alpha_p(D-3)!}{(D-2p-1)!}\left(\frac{\kappa-f(r)}{r^2%
}\right)^p=\frac{\mu}{r^{D-1}}\,,  \label{master}
\end{equation}
where $\mu$ is  an integration constant related to the mass of the solution, as we shall discuss  below. In case of asymptotically AdS spaces, $%
f(r)$ behaves as
\begin{equation}
f(r)=\kappa+\frac{r^{2}}{\ell_{\mathrm{eff}}^{2}}+\epsilon (r)\,,
\end{equation}
where $f_{\mathrm{AdS}}=\kappa +\frac{r^{2}}{\ell _{\mathrm{eff}}^{2}}$ is
the global AdS space and $\epsilon (r)$ is a function that, in most cases,
decays sufficiently fast. Plugging this into the master equation and
expanding in the asymptotic region ($1/r \rightarrow 0$), we find
\begin{eqnarray}
\frac{\mu }{r^{D-1}} &=&\sum_{p=0}^{N}\frac{\left( -1\right)
^{p}(D-3)!\alpha _{p}}{\left( D-2p-1\right) !}\,\left( \frac{1}{\ell_{%
\mathrm{eff}}^{2}} +\frac{\epsilon }{r^{2}}\right) ^{p}  \notag \\
&=&\sum_{q=0}^{N}\sum_{p=q}^{N}\frac{\left( -1\right) ^{p}(D-3)!\alpha _{p}}{%
\left( D-2p-1\right) !}\binom{p}{q}\left(\frac{1}{\ell_{\mathrm{eff}}^{2}}%
\right) ^{p-q}\left( \frac{\epsilon }{r^{2}}\right) ^{q}\,,
\end{eqnarray}
or equivalently
\begin{equation}
\frac{\mu }{r^{D-1}}=\sum_{q=0}^{N}\Delta ^{(q)}\,\left( \frac{\epsilon }{%
r^{2}}\right) ^{q}\,,
\end{equation}%
where $\Delta^{(q)}$ is given by eq.(\ref{DeltaK}). We can expand the above
series knowing that, for a $k$-fold degenerate vacuum, the first nonvanishing
term is $\Delta ^{(k)}$ yielding
\begin{equation}
\frac{\mu }{r^{D-1}}=\Delta ^{(k)}\,\left( \frac{\epsilon }{r^{2}}\right)
^{k}+\Delta ^{(k+1)}\,\left( \frac{\epsilon }{r^{2}}\right) ^{k+1}+\cdots \,.
\label{Expansion1}
\end{equation}
Furthermore, if $\Delta^{(k+1)}=0$, it can be shown that $\Delta^{(q)}=0$
for all $q>k$. {Then $\epsilon (r)$ can be solved exactly as $\epsilon
=\left( \frac{\mu }{\Delta ^{(k)}\,r^{D-2k-1}}\right)^{1/k}$.}

For an arbitrary value of $\Delta ^{(k+1)}$, we can assume an asymptotic
behavior for $\epsilon(r)$ of the form
\begin{equation}
\epsilon (r)=\frac{A}{r^{x}}+\frac{B}{r^{y}}+\cdots \,,\qquad y>x \geq
0\,,\qquad A\neq 0\,,  \label{e}
\end{equation}
where $A$, $B$, $x$ and $y$ are coefficients to be determined. Inserting eq.(%
\ref{e}) into (\ref{Expansion1}), we obtain
\begin{equation}
\frac{\mu }{r^{D-1}}=\Delta ^{(k)}\,\left( \frac{A^{k}}{r^{(x+2)k}}+\frac{%
kA^{k-1}B}{r^{(x+2) (k-1) +y+2}}+\cdots \right) +\Delta ^{(k+1)}\,\left(
\frac{A^{k+1}}{r^{(x+2)(k+1)}}+\cdots \right) .
\end{equation}
At leading order, we find
\begin{equation}
\frac{\mu }{r^{D-1}}=\Delta ^{(k)}\,\frac{A^{k}}{r^{\left( x+2\right) k}}\,,
\end{equation}
and therefore
\begin{equation}
x=\frac{D-2k-1}{k}\,,\qquad A=\left(\frac{\mu }{\Delta ^{(k)}}\right) ^{%
\frac{1}{k}}\,.  \label{x,A}
\end{equation}
Notice that the mass parameter may be negative and that $A$ can have any sign for even $k$. The subleading contributions
become
\begin{equation}
0=\Delta ^{(k)}\,\left( \frac{kA^{k-1}B}{r^{\left( x+2\right) \left(
k-1\right) +y+2}}+\cdots \right) +\Delta ^{(k+1)}\,\left( \frac{A^{k+1}}{%
r^{\left( x+2\right) \left( k+1\right) }}+\cdots \right) \,.
\label{sub-leading}
\end{equation}

We can distinguish three cases for the coefficients $A$ and $B$:

\begin{itemize}
\item[\textit{a})] When $A=B=0$, the solution becomes the vacuum state of
the theory, $\mu =0$, i.e., global AdS.

\item[\textit{b})] When $A\neq 0$ and $B=0$, the solution exists only if $%
\Delta ^{(k+1)}=0$. As discussed after eq.(\ref{Expansion1}), $\Delta ^{(k)}$
is the only nonvanishing coefficient and $\epsilon$ can be solved exactly.
This  case  is that of LUV theories.

\item[\textit{c})] When $A,B\neq 0$, the terms along $\Delta ^{(k)}$ and $%
\Delta ^{(k+1)}$ are of the same order. That implies
\begin{equation}
y=\frac{2 \left( D-1\right) -2k}{k}\,,  \label{y}
\end{equation}
and, in turn, the coefficient $B$ reads
\begin{equation}
B=-\frac{\Delta ^{(k+1)}A^{2}}{k\Delta ^{(k)}}=-\frac{\Delta ^{(k+1)}}{%
k\Delta ^{(k)}}\left( \frac{\mu }{\Delta ^{(k)}}\right) ^{\frac{2}{k}}.
\label{B}
\end{equation}
\end{itemize}

The cases \emph{a})--\emph{c}) are covered by the formulas (\ref{x,A}), (\ref%
{y}) and (\ref{B}). In this way, for a degenerate vacuum with multiplicity $%
k $, we find a general falloff of the metric function as
\begin{equation}
f(r)=\kappa+\frac{r^{2}}{\ell _{\mathrm{eff}}^{2}}+\left( \frac{\mu }{\Delta
^{(k)}r^{D-2k-1}}\right) ^{\frac{1}{k}} -\frac{\Delta ^{(k+1)}}{k {\Delta
^{(k)}}^{2}}\left( \frac{\mu^{2} }{\Delta ^{(k)}r^{2(D-1) -2k}}\right) ^{%
\frac{1}{k}}+\cdots \,.  \label{f(r) K}
\end{equation}

We illustrate the above relation with the exact solutions in LUV gravity
for $k<N$ \cite{Crisostomo:2000bb}
\begin{equation}
f_{\mathrm{LUV}}(r)=\kappa +\frac{r^{2}}{\ell _{\mathrm{eff}}^{2}}-\left(
\frac{2GM}{r^{D-2k-1}}\right) ^{\frac{1}{k}}\,,  \label{LUV-BH}
\end{equation}%
where $G$ is the gravitational constant. In maximally degenerate cases,
these are Chern-Simons and Born-Infeld AdS black holes
\begin{eqnarray}
f_{\mathrm{CS}}(r) &=&\kappa +\frac{r^{2}}{\ell _{\mathrm{eff}}^{2}}-(2GM+1)^{\frac{1}{n}}\,,  \notag \\
f_{\mathrm{BI}}(r) &=&\kappa +\frac{r^{2}}{\ell _{\mathrm{eff}}^{2}}-\left(
\frac{2GM}{r}\right) ^{\frac{1}{n-1}}\,.  \label{CS,BI-BH}
\end{eqnarray}
The mass parameter in  the Chern-Simons case has been redefined
so that the horizon
shrinks to a point when $M \rightarrow 0$. This produces a mass gap, $M=-1/2G $,
between the Chern-Simons black hole, $M\geq 0$, and global AdS space with $M=-1/2G$.

It is worthwhile noting that the identification of the mass parameter $M$ (related to the integration constant $\mu$ in the general formula (\ref{f(r) K})) as the total mass of the black hole was made in ref.\cite{Crisostomo:2000bb} and, for $k=2$, in ref.\cite{Fan:2016zfs}, based on   thermodynamic calculations. In \cite{Crisostomo:2000bb}, it has been also obtained as a Hamiltonian mass in the minisuperspace approach. In our method, however, we calculate the mass as the one that comes from Noether theorem, once the action has been supplemented by adequate boundary terms.  This notion agrees with the thermodynamic and Hamiltonian mass.

In what follows, we make extensive use of the falloff of the metric (\ref{f(r) K}) to find both the AdS curvature and the corresponding Weyl tensor.

\subsection{AdS curvature and Weyl tensor}

\label{riewf}

The only nonvanishing components of the Riemann tensor for the static
solution (\ref{ansatz}) are
\begin{equation}
\begin{array}{ll}
R_{tr}^{tr}=-\frac{1}{2}\,f^{\prime \prime }\,,\medskip \qquad &
R_{rm}^{rn}=R_{tm}^{tn}=-\frac{f^{\prime }}{2r}\,\delta _{m}^{n}\,, \\
R_{m_{1}m_{2}}^{n_{1}n_{2}}=\frac{\kappa-f}{r^{2}}\,\delta
_{m_{1}m_{2}}^{n_{1}n_{2}}\,. &
\end{array}%
\end{equation}
Using the falloff (\ref{f(r) K}), we get {%
\begin{eqnarray}
R_{tr}^{tr} &=&-\frac{1}{\ell _{\mathrm{eff}}^{2}}-\frac{\left(
D-2k-1\right) \left( D-k-1\right) }{2k^{2}}\left( \frac{\mu }{\Delta
^{(k)}r^{D-1}}\right) ^{\frac{1}{k}}+\mathcal{O}\left( r^{-\frac{2(D-1)}{k}%
}\right) \,,  \notag \\
R_{rm}^{rn} &=&R_{tm}^{tn}=\left[ -\frac{1}{\ell _{\mathrm{eff}}^{2}}+\frac{%
D-2k-1}{2k} \left( \frac{\mu }{\Delta ^{(k)}r^{D-1}}\right) ^{\frac{1}{k}}\right] \delta _{m}^{n}+\mathcal{O}\left( r^{-\frac{2(D-1)}{k}%
}\right) \,,  \notag \\
R_{m_{1}m_{2}}^{n_{1}n_{2}} &=&\left[ -\frac{1}{\ell _{\mathrm{eff}}^{2}}%
+\left( \frac{\mu }{\Delta ^{(k)}\,r^{D-1}}\right) ^{\frac{1}{k}}\right]
\delta _{m_{1}m_{2}}^{n_{1}n_{2}}+\mathcal{O}\left( r^{-\frac{2(D-1)}{k}%
}\right) \,.  \label{Riem}
\end{eqnarray}%
} Furthermore, the AdS curvature
\begin{equation}
F_{\alpha \beta }^{\mu \nu }=R_{\alpha \beta }^{\mu \nu } +\frac{1}{{\ell_{%
\mathrm{eff}}^{2}}}\,{\delta }_{\alpha \beta }^{\mu \nu }\,,
\label{AdScurvature}
\end{equation}
is the only part of the field strength of the local $SO(D-1,2)$ group that differs from zero in a Riemannian geometry. Using eqs.(\ref{Riem}), it
is straightforward to evaluate it as {%
\begin{eqnarray}
F_{tr}^{tr} &=&-\frac{(D-2k-1)(D-k-1)}{2k^2}\left( \frac{\mu }{\Delta
^{(k)}r^{D-1}}\right)^{\frac{1}{k}} +\mathcal{O}\left( r^{-\frac{2(D-1)}{k}%
}\right) \,,  \notag \\
F_{rm}^{rn} &=&F_{tm}^{tn}=\frac{D-2k-1}{2k}\left( \frac{\mu }{\Delta
^{(k)}r^{D-1}}\right)^{\frac{1}{k}}\, \delta_m^n+\mathcal{O}\left(r^{-%
\frac{2(D-1)}{k}}\right) \,,  \notag \\
F_{m_{1}m_{2}}^{n_{1}n_{2}} &=&\left( \frac{\mu }{\Delta ^{(k)}\,r^{D-1}}
\right) ^{\frac{1}{k}}\delta _{m_{1}m_{2}}^{n_{1}n_{2}} +\mathcal{O}%
\left(r^{-\frac{2(D-1)}{k}}\right) \,.  \label{AdSasymp}
\end{eqnarray}%
}

We can express the above in a useful way in terms of the Weyl tensor, which is defined in terms of the Riemann tensor and its contractions as
\begin{equation}
W_{\alpha \beta}^{ \mu \nu }=R_{ \alpha \beta }^{ \mu \nu }- \frac { 1 }{
D-2 } \,\delta_{ [\alpha }^{ [\mu }R_{ \beta ] }^{ \nu] } +\frac {R}{%
(D-1)(D-2) } \,{\delta }_{ \alpha \beta }^{\mu \nu }\,,
\end{equation}
{where the second term is the skew-symmetric product between the Kronecker
delta and the Ricci tensor, $\delta_{[\alpha}^{[\mu } R_{\beta ]}^{\nu ]}
=\delta _{\alpha }^{\mu }R_{\beta }^{\nu}-\delta _{\alpha }^{\nu }R_{\beta
}^{\mu } -\delta _{\beta }^{\mu }R_{\alpha}^{\nu }+\delta _{\beta }^{\nu
}R_{\alpha }^{\mu }$.}
In Einstein AdS gravity, the on-shell Weyl tensor coincides with the AdS
curvature. However in Lovelock AdS gravity the higher-order
contributions modify the above relation such that the on-shell Weyl tensor
can be written as
\begin{equation}
W_{\alpha \beta }^{\mu \nu }=F_{\alpha \beta }^{\mu \nu }+X_{\alpha \beta
}^{\mu \nu }\,,
\end{equation}%
where the tensor $X_{\alpha \beta }^{\mu \nu }$ is constructed from the
generalized Lanczos tensor (\ref{LL}) and its trace $H=H_{\mu }^{\mu }$,
\begin{equation}
X_{\alpha \beta }^{\mu \nu }=\left( \frac{1}{\ell ^{2}}-\frac{1}{\ell _{%
\mathrm{eff}}^{2}}+\frac{2H}{(D-1)(D-2)}\right) \delta _{\alpha \beta }^{\mu
\nu }-\frac{1}{D-2}\,{{\delta }_{[\alpha }^{[\mu }H_{\beta ]}^{\nu ]}}\,.
\label{X}
\end{equation}%
A computation based on the falloff of the Riemann tensor \eqref{Riem} computed above
shows that the components of $X$ are always subleading in $r$ with respect
to the AdS curvature. More precisely, $X_{\alpha \beta }^{\mu \nu }=\mathcal{%
O}\left( \mu ^{2/k}/r^{2(D-1)/k}\right) $ asymptotically. Therefore, the
leading order of the AdS curvature and the Weyl tensor is
\begin{equation}
W_{\alpha \beta }^{\mu \nu }=F_{\alpha \beta }^{\mu \nu }=\mathcal{O}\left(
\mu ^{1/k}/r^{(D-1)/k}\right) \,.  \label{WeylAsymp}
\end{equation}

In the next section we show that, for a given theory with degeneracy $k$,
the conserved charge formula is proportional to the $k$-th power of the Weyl
tensor. This is obtained as a consistent truncation of the charges found in
ref.\cite{kofinas,Kofinas:2008ub} in order to produce a finite energy flux
at the asymptotic region.

\section{Kounterterms and conserved charges}
\label{sec3}

The Lovelock AdS action is infrared (IR) divergent and has to be renormalized by adding
boundary counterterms. Instead of obtaining the local counterterm series
perturbatively, as in standard Holographic Renormalization \cite%
{Henningson:1998gx,Balasubramanian:1999re,Mann:1999pc,deHaro:2000vlm}, the idea behind the Kounterterm method
\cite{Olea:2005gb,Olea:2006vd} is that the bulk action is supplemented with
an appropriate boundary term that is linked either to topological
invariants or Chern-Simons forms. In $D=d+1$ dimensions, the renormalized
action defined on the manifold $\mathcal{M}$ reads
\begin{equation}
I_{\mathrm{ren}}={I}_{\mathrm{bulk}}+{c}_{d}\int\limits_{\partial {\mathcal{M%
}}}{{d}^{d}x\,{B}_{d}}(h,K,\mathcal{R})\,,
\label{Ireg}
\end{equation}
where $B_{d}(h,K,{\mathcal{R}})$ is a scalar density on the boundary $%
\partial \mathcal{M}$ that depends on the boundary metric, the extrinsic
curvature, and the boundary curvature. The overall factor ${c}_{d}$ is a
given coupling. It has been shown that the method appropriately leads to
finite conserved charges in higher-curvature gravity theories \cite{kofinas,
Giribet:2018hck} and it is also useful for computing holographic quantities
such as entanglement entropy \cite{Anastasiou:2018rla}.

As usual, the charges are expressed as an integral over a co-dimension 2
surface at fixed time and radial infinity. We can proceed taking a radial
foliation of the spacetime $\mathcal{M}$ in Gauss-normal coordinates,
\begin{equation}
ds^{2}=N^{2}(r)\,dr^{2}+h_{ij}(r,x)\,dx^{i}dx^{j}\,,
\end{equation}%
where $N(r)$ is the lapse function and $h_{ij}$ is the induced metric at a
fixed $r$. In turn, the boundary metric admits a time-like ADM foliation as
\begin{equation}
h_{ij}dx^{i}dx^{j}=-\tilde{N}^{2}(t)dt^{2}+\sigma _{mn}\left( dy^{m}+\tilde{N%
}^{m}dt\right) \left( dy^{n}+\tilde{N}^{n}dt\right) \,,
\end{equation}%
where now $\sqrt{-h}=\tilde{N}\sqrt{\sigma }$, with $\sigma _{mn}$ the
co-dimension two metric of the asymptotic boundary $\Sigma _{\infty }$. The
unit normal to the hypersurface is given by $u_{j}=(u_{t},u_{m})=(-\tilde{N}%
,0)$ and therefore, the conserved charges are given by the surface integral
\begin{equation}
Q[\xi ]=\int\limits_{\Sigma _{\infty }}{{d}^{d-1}y\sqrt{\sigma }\,{u}_{j}{%
\xi }^{i}}\left( \tau _{i}^{j}+\tau _{(0)i}^{j}\right) \,,  \label{Q}
\end{equation}%
where $\xi ^{i}$ is an asymptotic Killing vector. The charge density tensor
is naturally split in two contributions: $\tau _{i}^{j}$ that, when
integrated, can be identified with the mass and angular momentum of the
black hole, and $\tau _{(0)i}^{j}$ is associated with the vacuum/Casimir
energy in the context of AdS/CFT correspondence \cite{Balasubramanian:1999re}.

In even dimensions $D=2n$, $\tau _{i}^{j}$ has the form \cite{kofinas}
\begin{eqnarray}
\tau _{i}^{j} &=&\frac{1}{2^{n-2}}\,\delta _{{i}_{1}{i}_{2}\dots {i}_{2n-1}}^{j{j}_{2}\dots {j}_{2n-1}}\,K_{i}^{{i}_{1}}\left[ \frac{1}{16\pi
G}\sum_{p=1}^{n-1}{\frac{p{\alpha _{p}}}{(2n-2p)!}%
\,R_{j_{2}j_{3}}^{i_{2}i_{3}}\cdots
R_{j_{2p-2}j_{2p-1}}^{i_{2p-2}i_{2p-1}}\times }\right.  \notag \\
&&\qquad \qquad \times \left. \delta
_{j_{2p}j_{2p+1}}^{i_{2p}i_{2p+1}}\cdots \delta
_{j_{2n-2}j_{2n-1}}^{i_{2n-2}i_{2n-1}}+n{c}%
_{2n-1}R_{j_{2}j_{3}}^{i_{2}i_{3}}\cdots
R_{j_{2n-2}j_{2n-1}}^{i_{2n-2}i_{2n-1}}\rule{0pt}{17pt}\right] ,
\label{qqeven}
\end{eqnarray}%
and $\tau _{(0)i}^{j}=0$. The coupling in \eqref{Ireg} is fixed from the action principle,
\begin{equation}
c_{2n-1}=-\frac{1}{16\pi nG}\sum_{p=1}^{n-1}\frac{p\alpha _{p}}{(D-2p)!}%
\,(-\ell _{\mathrm{eff}}^{2})^{n-p}\,.
\end{equation}%

In odd dimensions $D=2n+1$, the charge density tensor
reads
\begin{eqnarray}
\tau _{i}^{j} &=&\frac{1}{2^{n-2}}\,\delta _{{i}_{1}{i}_{2}\dots {i}%
_{2n}}^{j{j}_{2}\dots {j}_{2n}}\,K_{i}^{{i}_{1}}{\delta }_{{j}_{2}}^{{i}%
_{2}}\left[ \frac{1}{16\pi G}\sum_{p=1}^{n}{\frac{p{\alpha _{p}}}{(2n-2p+1)!}%
R_{j_{3}j_{4}}^{i_{3}i_{4}}\cdots R_{j_{2p-1}j_{2p}}^{i_{2p-1}i_{2p}}\delta
_{j_{2p+1}j_{2p+2}}^{i_{2p+1}i_{2p+2}}}\cdots \right.  \notag \\
&&\hspace{-0.3cm}\left. \cdots \delta _{j_{2n-1}j_{2n}}^{i_{2n-1}i_{2n}}+n{c}%
_{2n}\int\limits_{0}^{1}{du\left( R_{j_{3}j_{4}}^{i_{3}i_{4}}+\frac{{u}^{2}}{%
\ell _{\mathrm{eff}}^{2}}\delta _{j_{3}j_{4}}^{i_{3}i_{4}}\right) \cdots }%
\left( R_{j_{2n-1}j_{2n}}^{i_{2n-1}i_{2n}}+\frac{{u}^{2}}{\ell _{\mathrm{eff}%
}^{2}}\delta _{j_{2n-1}j_{2n}}^{i_{2n-1}i_{2n}}\right) \right]  \label{qqodd}
\end{eqnarray}
and
\begin{equation*}
\tau _{(0)i}^{j}=-\frac{nc_{2n}}{2^{n-2}}\,\int_{0}^{1}du\,u\,\delta
_{{i}_{1}{ i}_{2}...{i}_{2n}}^{j{ j}_{2}...{ j}%
_{2n}}\left( {\delta }_{{ j}_{2}}^{{ i}_{2}}K_{i}^{{ i}_{1}}+{ \delta
}_{{ i}}^{{ i}_{2}}{ K}_{{ j}_{2}}^{{ i}_{1}}\right){ \mathcal{F}}_{{%
 j}_{3}{ j}_{4}}^{{ i}_{3}{i}_{4}}(u)\cdots {\mathcal{F}}_{{ j}%
_{2n-1}{ j}_{2n}}^{{ i}_{2n-1}{i}_{2n}}(u)\,,
\end{equation*}
where
\begin{equation}
\mathcal{F}_{lk}^{ij}(u)={\ R}_{lk}^{ij}-\left( u^{2}-1\right) \left(
K_{k}^{i}K_{l}^{j}-K_{l}^{i}K_{k}^{j}\right)+\frac{u^{2}}{\ell _{\mathrm{eff}}^{2}}{\ \delta }_{kl}^{ij}\,.
\end{equation}%
The coupling constant in this case is
\begin{equation}
c_{2n}=-\frac{1}{16\pi nG}\left[ \int\limits_{0}^{1}du(1-u^{2})^{n-1}\right]
^{-1}\sum_{p=1}^{n}\frac{p\alpha _{p}}{(D-2p)!}\,(-\ell _{\mathrm{eff}%
}^{2})^{n-p}\,.  \label{c_odd}
\end{equation}

The charge in $D$ dimensions is a polynomial of order $N$ in the curvature.
In nondegenerate Lovelock theories, this polynomial can be truncated so that
it becomes linear in the Weyl tensor \cite{Arenas-Henriquez:2017xnr}
\begin{equation}
\tau _{i}^{j}=-\frac { { { \ell _{ \rm{ eff } }\, \Delta ^{ (1) }(\ell_{\rm{eff}}^{-2}) } } }{ 32\pi G\, (D-3) } \, \delta _{ ii_{ 2 }{ i }_{ 3 } }^{ jj_{ 2 }{ j }_{ 3 } }\, { W }_{ { j }_{ 2 }{ j }_{ 3 } }^{ { i }_{ 2 }{ i }_{ 3 } } \,.
\end{equation}
This expression corresponds to Conformal Mass for nondegenerate Lovelock AdS gravity \cite%
{Arenas-Henriquez:2017xnr}, as an extension of the concept developed by
Ashtekar, Magnon and Das \cite{Ashtekar:1984zz, Ashtekar:1999jx}. Using the identity
$\delta _{ii_{2}}^{jj_{2}}{W}_{{j}_{2}{j}_{3}}^{{i}_{2}{i}_{3}}=4W_{ri}^{rj}$
the charge becomes proportional to  the electric part of the Weyl tensor
\begin{equation}
\tau _{i}^{j}=-\frac{\ell_{\rm{eff}}}{8\pi G} \, \Delta^{(1)}(\ell_{\rm{eff}}^{-2})E^{j}_{i}\,.
\label{conformalmass}
\end{equation}
where   in $D$ dimensions
\begin{equation}
E^{j}_{i}=\frac{1}{D-3}\,W^{rj}_{ri}\, .
\end{equation}
Clearly \eqref{conformalmass} fails when $\Delta^{(1)}=0$.

When
the theory has multiplicity $k$, a power-counting argument indicates that the
asymptotic falloff of $\tau_{i}^{j}$ is such that the system has finite global
charges. Namely, the bulk metric behaves as in eq.(\ref{f(r) K}), and so $%
\sqrt{\sigma}= \mathcal{O}(r^{D-2})$, $u_{j} = \mathcal{O}(r)$ and $\xi =
\mathcal{O}(1)$. Since the charge $Q[\xi]$ is of order $\mathcal{O}(1)$, it
implies that $\tau _{i}^{j}$ should be of order $\mathcal{O}(1/r^{D-1})$. At
the same time, the charge should be zero for global AdS. Hence it must be a nonlinear expression in the Weyl tensor.

\section{Generalized Conformal Mass}
\label{sec4}

In this section we manipulate the general formulas (\ref{qqeven}) and (\ref%
{qqodd}) for $\tau_{i}^{j}$ to make manifest the dependence on the
degeneracy conditions at different orders.

\subsection{Even dimensions $D=2n$}

The charge density in even dimensions can be conveniently rewritten as
\begin{eqnarray}
\tau _{i}^{j} &=&\frac{{{\ell _{\mathrm{eff}}^{2n-2}}}}{16\pi G\, {2}^{n-2}}%
\,\delta _{i_{1}\cdots i_{2n-1}}^{jj_{2}\cdots
j_{2n-1}}\,K_{i}^{i_{1}}\sum_{p=1}^{n-1}{\frac{p{\alpha }_{p}}{(2n-2p)!}}%
\left( \frac{1}{{{\ell _{\mathrm{eff}}^{2}}}}\right)
^{p-1}R_{j_{2}j_{3}}^{i_{2}i_{3}}\cdots
R_{j_{2p-2}j_{2p-1}}^{i_{2p-2}i_{2p-1}}\times  \notag \\
&&\hspace{-0.3cm}\left[ {\left( \frac{1}{{{\ell _{\mathrm{eff}}^{2}}}}%
\right) ^{n-p}\delta _{j_{2p}j_{2p+1}}^{i_{2p}i_{2p+1}}}\cdots \delta
_{j_{2n-2}j_{2n-1}}^{i_{2n-2}i_{2n-1}}-{(-1)}%
^{n-p}R_{j_{2p}j_{2p+1}}^{i_{2p}i_{2p+1}}\cdots
R_{j_{2n-2}j_{2n-1}}^{i_{2n-2}i_{2n-1}}\right] . \label{echarge}
\end{eqnarray}%
As it is clear from the above formula, the charge is a polynomial of order $%
n-1$ in the curvature. For a branch with the degeneracy $k$, this can be
rearranged in order to express the polynomial as a product of $k$ AdS
curvatures times a polynomial $\mathcal{P}(R)$ of order $n-1-k$ in the
curvature
\begin{align}
\tau _{i}^{j}& =\frac{{{\ell _{\mathrm{eff}}^{2(n-1)}}}}{16\pi G\,{2}^{n-2}}\,\delta _{i_{1}i_{2}\cdots {i}_{2k}{i}_{2k+1}\dots {i}_{2n}i_{2n-1}}^{jj_{2}\cdots {j}_{2k}{j}_{2k+1}\dots {j}_{2n}j_{2n-1}}K_{i}^{i_{1}}\left( R_{j_{2}j_{3}}^{i_{2}i_{3}}+\frac{1}{\ell_{\mathrm{eff}}^{2}}\delta _{j_{2}j_{3}}^{i_{2}i_{3}}\right) \times \dots
\notag \\
& \qquad\qquad \dots \times \left( R_{j_{2k}j_{2k+1}}^{i_{2k}i_{2k+1}}+\frac{1}{\ell _{\mathrm{eff}}^{2}}\delta _{j_{2k}j_{2k+1}}^{i_{2k}i_{2k+1}}\right){\mathcal{P}_{j_{2k+2}\cdots j_{2n-1}}^{i_{2k+2}\cdots i_{2n-1}}}(R)\,,
\label{goal}
\end{align}
where $\mathcal{P}(R)$ written as in eq.(\ref{P final}) of Appendix \ref{Fact} with all the indices
\begin{eqnarray}
&&\left. \mathcal{P}{_{j_{2k+2}\cdots j_{2n-1}}^{i_{2k+2}\cdots i_{2n-1}}}%
(R)=\sum_{s=0}^{n-k-1}\sum_{p=k}^{k+s}\Delta ^{(p)}(C_{k+s,p}){%
_{j_{2k+2}\cdots j_{2n-2s-1}}^{i_{2k+2}\cdots i_{2n-2s-1}}\times }\right.
\notag \\
&&\hspace{-0.3cm}\left( R_{j_{2n-2s}j_{2n-2s+1}}^{i_{2n-2s}i_{2n-2s+1}}+%
\frac{1}{\ell _{\mathrm{eff}}^{2}}\delta
_{j_{2n-2s}j_{2n-2s+1}}^{i_{2n-2s}i_{2n-2s+1}}\right) \cdots \left(
R_{j_{2n-2}j_{2n-1}}^{i_{2n-2}i_{2n-1}}+\frac{1}{\ell _{\mathrm{eff}}^{2}}%
\delta _{j_{2n-2}j_{2n-1}}^{i_{2n-2}i_{2n-1}}\right)  \label{P}
\end{eqnarray}
For a detailed construction of this factorization in even dimensions, see Appendix \ref{evenfactorization}.
The charge has to be evaluated in the asymptotic region where we know that
the curvature tensor behaves as in eq.(\ref{Riem}). In turn, the extrinsic
curvature has the asymptotic form
\begin{equation}
K_{j}^{i}=-\frac{1}{\ell _{\mathrm{eff}}}\,\delta _{j}^{i}+\mathcal{O}%
(1/r^{2})\,.  \label{extrinsic}
\end{equation}%
Furthermore, given the fact that the charge density at large distances is $\tau _{j}^{i}=\mathcal{O}(1/r^{D-1})$,  the AdS curvature falloff (\ref{AdSasymp}) constrains the polynomial $\mathcal{P}=\mathcal{O}(1)$ to  leading order. This implies that the Riemann curvature
is $R=-\frac{1}{\ell _{\mathrm{eff}}^{2}}\delta^{[2]}$ in the
polynomial so that the only nonzero contribution in eq.(\ref{P}) comes from $s=0,p=k$, corresponding to the coefficient $C_{kk}$ given by
\begin{equation}
(C_{kk}){_{j_{2k+2}\cdots j_{2n-1}}^{i_{2k+2}\cdots i_{2n-1}}}=\frac{%
(-1)^{k-1}}{2(2n-3)!\ell _{\mathrm{eff}}^{2(n-2)}}\,\delta {%
_{j_{2k+2}j_{2k+3}}^{i_{2k+2}i_{2k+3}}}\cdots \delta {_{j_{2n-2}\cdots
j_{2n-1}}^{i_{2n-2}i_{2n-1}}}
\end{equation}
from eq.(\ref{cqq}) in Appendix \ref{evenfactorization}.
The above reasoning yields
\begin{equation}
\mathcal{P}_{j_{2k+2}\cdots j_{2n-1}}^{i_{2k+2}\cdots i_{2n-1}}(-\ell_{\mathrm{eff}}^{-2}\delta^{[2]})=-\frac{(-1)^{k}\Delta^{(k)}}{2(2n-3)!\ell_{\rm{eff}}^{2(n-2)}}\,\delta_{{j}_{2k+2}{j}%
_{2k+3}}^{{i}_{2k+2}{i}_{2k+3}}\cdots {\delta }_{{j}_{2n}{j}_{2n-1}}^{{i}%
_{2n}{i}_{2n-1}}.
\end{equation}%
Clearly, this quantity is nonvanishing only in a theory whose vacuum has
multiplicity $k$. Replacing the asymptotic form of the extrinsic curvature
of eq.(\ref{extrinsic}) and the relation between the AdS curvature and the
Weyl tensor of eq.(\ref{WeylAsymp}), we find that the charge density tensor in even dimensions can be consistently truncated up to the order $k$ in the Weyl tensor,
\begin{eqnarray}
\tau _i^j &=&\frac{\ell_{\mathrm{eff}}(-1)^k\,\Delta ^{(k)}}{16\pi G\,{2}^{n-1}(2n-3)!}\,\delta _{ii_{2}\cdots {i}_{2k}{i}_{2k+1}\dots {i}%
_{2n}i_{2n-1}}^{jj_{2}\cdots {j}_{2k}{j}_{2k+1}\dots {j}_{2n}j_{2n-1}}\,{W}_{%
{j}_{2}{j}_{3}}^{{i}_{2}{i}_{3}}\cdots {W}_{{j}_{2k}{j}_{2k+1}}^{{i}_{2k}{i}%
_{2k+1}}\times  \notag \\
&&\qquad \qquad \qquad \qquad \qquad \times {\delta }_{{j}_{2k+2}{j}%
_{2k+3}}^{{i}_{2k+2}{i}_{2k+3}}\cdots {\delta }_{{j}_{2n}{j}_{2n-1}}^{{i}%
_{2n}{i}_{2n-1}}\,.
\end{eqnarray}%
Upon a suitable contraction of the Kronecker deltas, the charge becomes
\begin{equation}
\tau _{i}^{j}=\frac{{{\ell_{\mathrm{eff}}{ (-1)}^{k}\,\Delta
^{(k)}(2n-2k-2)!}}}{16\pi G\,{2}^{k}(2n-3)!}\,\delta _{ii_{2}\cdots {i}%
_{2k}{ i}_{2k+1}}^{jj_{2}\cdots {j}_{2k}{ j}_{2k+1}}\,{W}_{{ j}_{2}{
j}_{3}}^{{ i}_{2}{\ i}_{3}}\cdots {W}_{{ j}_{2k}{j}_{2k+1}}^{{ i}_{2k}%
{ i}_{2k+1}}\,.
\end{equation}%
For $k=1$, the last formula reduces to the Conformal Mass \eqref{conformalmass} in nondegenerate Lovelock theories.
In the maximally degenerate case, i.e., Born-Infeld AdS gravity, the charge
is a product of $n-1$ Weyl tensors or AdS curvatures, which matches with the
result in ref.\cite{Miskovic:2007mg}. The degeneracy condition (\ref{DeltaK}%
) as an overall factor shows the validity of the formula for a particular
gravity theory with multiplicity $k$. Indeed, for theories with the
intermediate degeneracy $1<k<n-1$, we can drop $n-k-1$ curvatures from the
original charge formula (\ref{qqeven}).

\subsection{Odd dimensions $D=2n+1$}

\label{Odd+CS} In odd dimensions, black hole mass is associated with $%
\tau^{j}_{i}$ given by eq.(\ref{qqodd}), which can be recast as
\begin{equation}
\tau _i^j =\frac{(2n-1)!(-\ell _{\mathrm{eff}}^2)^{n-1}}{2^{3n-4}(n-1)!^2
16\pi G}\, \delta _{ i_{ 1 }\cdots i_{ 2n } }^{ jj_{ 2 }\cdots j_{ 2n } }K_{
i }^{ i_{ 1 } }\delta _{ j_{ 2 } }^{ i_{ 2 } }\sum _{ p=1 }^{ n } \frac {
(-1)^{ p }p\alpha _{ p }\, \ell _{ \mathrm{{eff } }}^{ 2(1-p) } }{
(2n-2p+1)! } \int _{ 0 }^{ 1 } du\, (\mathcal{I}_{p})_{ j_{ 3 }\cdots j_{
2n } }^{ i_{ 3 }\cdots i_{ 2n } }(u)\,,  \label{qodd}
\end{equation}
where the tensorial quantity $\mathcal{I}_{p}(u)$ is a polynomial of order $%
n-1$ in the curvature,
\begin{equation}
\mathcal{I}_{p}(u)=\left( R+\frac{u^{2}}{\ell _{\mathrm{eff}}^{2}}\,\delta
^{\lbrack 2]}\right) ^{n-1}-(u^2-1) ^{n-1}(-R)^{p-1}\left( \frac{1}{\ell _{%
\mathrm{eff}}^{2}}\,\delta ^{\lbrack 2]}\right) ^{n-p}.  \label{I}
\end{equation}

We will restrict our analysis to the theories with $k<n$. We can rearrange
the expression for $\tau_{i}^{j}$ and factorize it in a similar fashion as
in the even dimensional case (see Appendix \ref{oddfactorization}),
\begin{eqnarray}
\tau _i^j&=&\frac{ 2^{2n-2}{(n-1)!}^2\ell _{ \mathrm{eff}}^{ 2(n-1) } }{ 2^{ n-2 }(2n-1)! 16\pi G}\,\delta _{ i_{ 1 }i_{ 2 }\cdots {%
 i }_{ 2k }{ i }_{ 2k+1 }\dots { i }_{ 2n }i_{ 2n-1 } }^{ jj_{ 2 }\cdots {%
 j }_{ 2k }{j }_{ 2k+1 }\dots { j }_{ 2n }j_{ 2n-1 } }K_{ i }^{ i_{ 1 }
}\left( R_{ j_{ 2 }j_{ 3 } }^{ i_{ 2 }i_{ 3 } }+\frac { 1 }{ \ell _{ \mathrm{eff}}^2}\, \delta _{ j_{ 2 }j_{ 3 } }^{ i_{ 2 }i_{ 3 } } \right)
\times \dots  \notag \\
&& \qquad \qquad \dots \times \left( R_{j_{ 2k }j_{ 2k+1 } }^{i_{ 2k }i_{
2k+1 } }+\frac { 1 }{ \ell _{\mathrm{{eff } }}^{ 2 } }\, \delta _{j_{ 2k
}j_{2k+1 } }^{i_{ 2k }i_{ 2k+1 } } \right) \mathcal{ P }_{j_{
2k+2 }\cdots j_{ 2n-1 } }^{ i_{2k+2 }\cdots i_{2n-1 } }(R)\,,
\label{qoddfact}
\end{eqnarray}
where $\mathcal{P}(R)$ is a polynomial in the curvature tensor of
order $n-k-1$. It turns out that, again, it has the form of a linear combination of
the derivatives $\Delta ^{(p)}$ of the characteristic polynomial (see eq.(%
\ref{P final}) in Appendix \ref{Fact})
\begin{equation}
\mathcal{P}_{j_{2k+2}\cdots j_{2n-1}}^{i_{2k+2}\cdots
i_{2n-1}}(R)=\sum_{s=0}^{n-1-k}\sum_{p=k}^{k+s}\Delta
^{(p)}(C_{k+s,p})_{j_{2k+2}\cdots j_{2n-1}}^{i_{2k+2}\cdots
i_{2n-1}}F_{j_{2n-2}j_{2n-1}}^{i_{2n-2}i_{2n-1}}\cdots
F_{j_{2n-2}j_{2n-1}}^{i_{2n-2}i_{2n-1}}\,,
\end{equation}
with some tensorial coefficients $C_{k+s,p}$ that differ from the even dimensional case. Evaluating the polynomial in the
asymptotic region, the only part relevant for the conserved charge is
\begin{equation}
\mathcal{P}_{j_{2k+2}\cdots j_{2n-1}}^{i_{2k+2}\cdots i_{2n-1}}(-\ell _{%
\mathrm{eff}}^{-2}\delta ^{[2]})=\Delta
^{(k)}(C_{kk})_{j_{2k+2}\cdots j_{2n-1}}^{i_{2k+2}\cdots i_{2n-1}}
\end{equation}
where the
coefficient has the form
\begin{equation}
(C_{kk})_{j_{2k+2}\cdots j_{2n-1}}^{i_{2k+2}\cdots i_{2n-1}}=\frac{\left(
-1\right) ^{n-k-1}2^{2n-3}\left( n-1\right) !^{2}}{(2n-1)!}\,\ell _{\mathrm{%
eff}}^{2}\,{\delta }_{{j}_{2k+2}{j}_{2k+3}}^{{i}_{2k+2}{i}_{2k+3}}\cdots {%
\delta }_{{j}_{2n}{j}_{2n-1}}^{{i}_{2n}{i}_{2n-1}}
\end{equation}
as shown in eq.(\ref{cqq-odd}) of Appendix \ref{oddfactorization}.
Knowing the behavior of the extrinsic curvature (\ref{extrinsic}) and the
relation (\ref{WeylAsymp}), we can evaluate the charge asymptotically as
\begin{eqnarray}
\tau _{i}^{j} &=&\frac{{{\ell _{\mathrm{eff}}{(-1)}^{k}\Delta ^{(k)}}}}{%
16\pi G\,{2}^{n-1}(2n-2)!}\,\delta _{ii_{2}\cdots {i}_{2k}{i}_{2k+1}\dots {i}%
_{2n}i_{2n-1}}^{jj_{2}\cdots {j}_{2k}{j}_{2k+1}\dots {j}_{2n}j_{2n-1}}\,{W}_{%
{j}_{2}{j}_{3}}^{{i}_{2}{i}_{3}}\cdots {W}_{{j}_{2k}{j}_{2k+1}}^{{i}_{2k}{i}%
_{2k+1}}\times  \notag \\
&&\qquad \qquad \qquad\qquad \qquad \times {\delta }_{{j}_{2k+2}{j}_{2k+3}}^{%
{i}_{2k+2}{i}_{2k+3}}\cdots {\delta }_{{j}_{2n}{j}_{2n-1}}^{{i}_{2n}{i}%
_{2n-1}}\,.
\end{eqnarray}
Finally, contracting the antisymmetric deltas, the charge density becomes
\begin{equation}
\tau ^{ j }_{ i }=\frac { { { \ell _{ \mathrm{{\ eff } }}{\ (-1) }^{ k
}\Delta ^{ (k) }(2n-2k-1)! } } }{ 16\pi G\, {\ 2 }^{ k }(2n-2)! } \, \delta
_{ii_{ 2 }\cdots {i }_{ 2k }{ i }_{ 2k+1 } }^{ jj_{ 2 }\cdots {j }_{
2k }{ j }_{ 2k+1 } }\, {W }_{{ j }_{ 2 }{j }_{ 3 } }^{ { i }_{ 2 }{
i }_{ 3 } }\cdots {W }_{ {j }_{ 2k }{j }_{ 2k+1 } }^{ { i }_{ 2k }{
i }_{ 2k+1 } }\,.
\end{equation}
The last expression is valid for theories whose AdS vacua have a degeneracy
level in the interval $1\leq k \leq n-1$. It is a generalization of the
Ashtekar-Magnon-Das conformal mass formula \cite{Ashtekar:1984zz,Ashtekar:1999jx} to the Lovelock AdS gravity \cite%
{Arenas-Henriquez:2017xnr}. Once again, for $k=1$ it reduces to the known
Conformal Mass (\ref{conformalmass}).

In the case of Chern-Simons AdS gravity, the mass term does not fall off as $%
r\rightarrow \infty$. The fact that the energy for global AdS is not
continuously connected with the spectrum of black holes of the theory
indicates that the energy for the vacuum state cannot be achieved by charge
that is proportional to the Weyl tensor. This presents a qualitative
difference between the theories with degeneracy $k<n$ and Chern-Simons AdS
gravity, which is reflected in their holographic properties \cite%
{Banados:2004zt,Banados:2005rz}.

\subsection{Energy of the LUV AdS black hole }

As an example, we calculate the total energy of the LUV AdS black hole
solution described by the metric (\ref{ansatz}) with the metric function (%
\ref{LUV-BH}). We can identify $\frac{\mu }{\Delta ^{(k)}}=2GM$. Because the
tensor $X_{\alpha \beta }^{\mu \nu }$ given by (\ref{X}) identically
vanishes in this case, the Weyl tensor and the AdS curvature are equal
on-shell, ${W}_{\alpha \beta }^{\mu \nu }={F}_{\alpha \beta }^{\mu \nu }$.
The degeneracy (\ref{DeltaK}) for the LUV gravity becomes
\begin{equation}
\Delta _{\mathrm{LUV}}^{(k)}=2^{k-1}(-1)^{k-1}(D-3)!\,.
\end{equation}

The total energy of the system is the Noether charge (\ref{Q}) for the time
translations $\xi =\partial _{t}$ and the boundary with the unit vector $%
u_{j}=-\sqrt{f}\,\delta _{j}^{t}$. Furthermore, we have $\sigma
_{nm}=r^{2}\gamma _{nm}(y)$, where the transverse metric $\gamma _{nm}$
depends only on the transverse coordinates $y^{m}$. Then we can find the
Jacobian ${\sqrt{\sigma }=r}^{D-2}\sqrt{\gamma }$ and the volume of the
transverse section $\Omega _{D-2}=\int\limits_{\Sigma _{\infty }}{{d}^{D-2}y%
\sqrt{\gamma }}$.

With this at hand, we can evaluate the total energy as%
\begin{equation}
E=Q[\partial _{t}]=E_{\mathrm{vacum}}-\Omega _{D-2}\left( {r}^{D-2}\sqrt{f}%
\tau _{t}^{t}\right) _{r\rightarrow \infty }\,,
\end{equation}%
where the vacuum energy for Lovelock gravity (the energy of the global AdS
space) exists in odd dimensions $D$ only, and was calculated in ref.\cite%
{kofinas}.

On the other hand, the charge density tensor $\tau _{i}^{j}$ given by eqs.(%
\ref{Teven}) and\ (\ref{Todd}) in even and odd dimensions, respectively, has
the form
\begin{equation}
\tau _{t}^{t}=\frac{{{{(-1)}^{k}}}\ell _{\mathrm{eff}}{{{\ }\Delta
^{(k)}(D-2k-2)!}}}{16\pi G\,{2}^{k}(D-3)!}\,\delta _{n_{1}\cdots {n}%
_{2k}}^{m_{1}\cdots {m}_{2k}}\,{W}_{{m}_{1}{m}_{2}}^{{n}_{1}{n}_{2}}\cdots {W%
}_{{m}_{2k-1}{m}_{2k}}^{{n}_{2k-1}{n}_{2k}}\,.
\end{equation}%
The only relevant components of the Weyl tensor read {%
\begin{equation}
W_{m_{1}m_{2}}^{n_{1}n_{2}}=\left( \frac{2GM}{r^{D-1}}\right) ^{\frac{1}{k}%
}\delta _{m_{1}m_{2}}^{n_{1}n_{2}}\,.  \label{WeylLUV}
\end{equation}%
Using the expression}%
\begin{equation}
\delta _{n_{1}\cdots {n}_{2k}}^{m_{1}\cdots {m}_{2k}}\,{W}_{{m}_{1}{m}_{2}}^{%
{n}_{1}{n}_{2}}\cdots {W}_{{m}_{2k-1}{m}_{2k}}^{{n}_{2k-1}{n}_{2k}}=\frac{2GM%
}{r^{D-1}}\,\frac{2^{k}\left( D-2\right) !}{\left( D-2k-2\right) !}\,,
\end{equation}%
it is straightforward to evaluate the energy density as
\begin{equation}
\tau _{t}^{t}=-\frac{\ell _{\mathrm{eff}}{\ }2^{k}\left( D-2\right) !}{32\pi
G\,}\,\frac{2GM}{r^{D-1}}\,,
\end{equation}%
and the total energy of the system%
\begin{equation}
E=E_{\mathrm{vacuum}}+\frac{\left( D-2\right) !2^{k}\Omega _{D-2}}{16\pi \,}%
\,M\,.  \label{E}
\end{equation}
The total mass, $E-E_{\mathrm{vacuum}}$, is indeed linear in the parameter $M$. Furthermore, in order to have the charge that is directly $M$ as in ref.\cite{Crisostomo:2000bb}, the gravitational action (\ref{action}), (\ref{alpha_LUV}) has to be normalized suitably, by dividing it by $\Omega _{D-2}$, a $D$-dependent factor and introducing the gravitational constant $G_{k}$ which explicitly depends on the multiplicity \cite{Footnote}. 
General Relativity ($k=1$) and BI gravity ($k=n-1$) are particular cases of LUV gravity in AdS space.

\section{Conclusions}
\label{sec5}

We have derived the Conformal Mass formula for a branch of Lovelock AdS gravity with
a $k$-fold vacuum degeneracy.  This comes as the generalization of the results found in ref.\cite{Arenas-Henriquez:2017xnr}, which indicate that the energy of
 black holes in that theory cannot be written as a linear expression in the AdS curvature (\ref{AdScurvature}) or, equivalently, in
 terms of the electric part of the Weyl tensor. On the contrary, we find that the conserved quantity must be nonlinear in the Weyl tensor in order
 to capture the falloff properties of the mass term in the metric.

 To summarize, the charge density tensor  has the form
\begin{equation}
\tau _{i}^{j}=\frac{{{{(-1)}^{k}}}\ell _{\mathrm{eff}}{{{\ }\Delta
^{(k)}(D-2k-2)!}}}{16\pi G\,{2}^{k}(D-3)!}\,\delta _{ii_{2}\cdots {i}_{2k}{i}%
_{2k+1}}^{jj_{2}\cdots {j}_{2k}{j}_{2k+1}}\,{W}_{{j}_{2}{j}_{3}}^{{i}_{2}{i}%
_{3}}\cdots {W}_{{j}_{2k}{j}_{2k+1}}^{{i}_{2k}{i}_{2k+1}}
\end{equation}
 in both even and odd dimensions, respectively, given by  eqs.(\ref{Teven}) and (\ref{Todd}).
As an example,  we showed that this formula gives the total energy of a static topological black hole in AdS space of  multiplicity $k$.

When $k=1$, this tensor becomes proportional to the electric part of the Weyl
tensor (\ref{conformalmass}). In general, when $k>1$, the
appearance of the degeneracy condition in the corresponding surface terms coming from an arbitrary variation of the
renormalized action (\ref{Ireg}) may be useful in understanding holographic properties of  degenerate AdS gravity theories. In
particular, it has been claimed \cite{Camanho:2013pda,Bueno:2018yzo} that the $C_{T}$ coefficient in holographic two-point functions is proportional to the first degeneracy
condition (\ref{DeltaPrime}). As the coefficient is linked to the
$a$-charge (type $A$-anomaly), recent results suggest that it is necessary to consider higher degeneracy conditions \cite{Li:2018drw} when one deals with theories with degenerate AdS vacua.

\section*{Acknowledgments}
G.A.H. and R.O. thank to G. Anastasiou, I.J. Araya, C. Arias, P. Bueno, F. Diaz and D. Rivera-Betancour for insightful comments.
This work was supported in parts by Chilean FONDECYT projects N$^{\circ}$1170765 ``Boundary dynamics in anti-de Sitter gravity and gauge/gravity
duality'' and N$^{\circ}$1190533 ``Black holes and asymptotic symmetries'' and in part by
the Natural Sciences and Engineering Research Council of Canada.

\appendix

\section{Factorization of the charge polynomial}
\label{Fact}

In this section we use the following shorthand notation. $\delta ^{\lbrack
p]}$ corresponds to the antisymmetric Kronecker delta of rank $p$, which is,
$\delta _{i_{1}\cdots i_{p}}^{j_{1}\cdots j_{p}}$, and $\delta
_{i[p]}^{j[p]} $ is a totally antisymmetric Kronecker delta of rank $p$ with
the indices $i$ and $j$ fixed, what means $\delta _{ii_{2}\cdots
i_{p}}^{jj_{2}\cdots j_{p}}$. Writing out the contracted indices will be
omitted.

In both even $D=2n$ and odd $D=2n+1$ dimensions, the charge density tensor
has the form
\begin{equation}
\tau _{i}^{j}=\frac{{\theta }_{D}}{16\pi G}\,\delta _{i_{1}\cdots
i_{2n-1}}^{jj_{2}\cdots j_{2n-1}}K_{i}^{i_{1}}f_{j_{2}\cdots
j_{2n-1}}^{i_{2}\cdots i_{2n-1}}(R)\,,
\end{equation}
or shortly
\begin{equation}
\tau _{j}^{i}=\frac{{\theta }_{D}}{16\pi G}\,\delta ^{\lbrack
2n-1]j}\,K_{i}\,f(R)\,,  \label{tau}
\end{equation}%
where the expressions for the particular coefficient ${\theta }_{D}$ and the
auxiliary function $f(R)$, which is the polynomial of order $n-1$ the
Riemann curvature, depends on the regularization method, as given by eqs.(%
\ref{qqeven}) and (\ref{qqodd}), respectively.

Let the Lovelock coefficients $\alpha _{p}$ are such that the theory has a
degenerate AdS vacuum of order $k$ given through the criterion (\ref{DeltaK}).
Then our main goal is to show that the function $f$ can be factorized by
$\left( R+\frac{1}{{\ell _{\mathrm{eff}}^{2}}}\,\delta ^{[2]}\right)^{k}$,
multiplied by a polynomial $\mathcal{P}$ of the order $n-k-1$ in the Riemann
curvature, which is \textit{finite} in the asymptotic region, and moreover
it is proportional to the degeneracy factor $\Delta ^{(k)}$.

Thus, we will prove that, in any dimension, the auxiliary function can be
put into the form%
\begin{equation}
f_{j_{2}\cdots j_{2n-1}}^{i_{2}\cdots i_{2n-1}}(R)=\left(
R_{j_{2}j_{3}}^{i_{2}i_{3}}+\frac{1}{{\ell _{\mathrm{eff}}^{2}}}\,\delta
_{j_{2}j_{3}}^{i_{2}i_{3}}\right) \cdots \left(
R_{j_{2k}j_{2k+1}}^{i_{2k}i_{2k+1}}+\frac{1}{{\ell _{\mathrm{eff}}^{2}}}%
\,\delta _{j_{2k}j_{2k+1}}^{i_{2k}i_{2k+1}}\right) \mathcal{P}%
_{j_{2k+2}\cdots j_{2n-1}}^{i_{2k+2}\cdots i_{2n-1}}(R)\,.
\end{equation}
Mathematically, these statements are easier formulated if we write $%
A_{kl}^{ij}=\frac{1}{{\ell _{\mathrm{eff}}^{2}}}\,\delta _{kl}^{ij}$, or
symbolically%
\begin{equation}
A=\frac{1}{{\ell _{\mathrm{eff}}^{2}}}\,\delta ^{\lbrack 2]}\,,  \label{a,b}
\end{equation}%
so that we have to prove%
\begin{equation}
f=(A+R)^{k}\mathcal{P}(R)\,,  \label{factor}
\end{equation}%
where the polynomial $\mathcal{P}(R)$ of the order $n-k-1$ cannot be further
factorized by $A+R$. Furthermore, $\mathcal{P}$ is finite when $R=-A$,
taking the value
\begin{equation}
\mathcal{P}(-A)=C_{kk}\,\Delta ^{(k)}\neq 0\,.
\end{equation}%
Because the tensor $f_{j_{2}\cdots j_{2n-1}}^{i_{2}\cdots i_{2n-1}}$ has $%
2n-2$ pairs of indices, and the tensor $\mathcal{P}_{j_{2k+2}\cdots
j_{2n-1}}^{i_{2k+2}\cdots i_{2n-1}}$ has $2n-2-2k$ pairs of indices, it
implies that $C_{kk}$ has the same index structure as $\mathcal{P}$, that
is, $\left( C_{kk}\right) _{j_{2k+2}\cdots j_{2n-1}}^{i_{2k+2}\cdots
i_{2n-1}}$.

The degeneracy condition (\ref{DeltaK}) can be written in a simpler way as%
\begin{equation}
\Delta ^{(q)}=\sum_{p=q}^{n-1}\binom{p}{q}\ell _{\mathrm{eff}}^{2(q-p)}\beta
_{p}\,,  \label{Delta-q}
\end{equation}%
where the new coefficients $\beta _{p}$ collect, along the Lovelock couplings $%
\alpha _{p}$, also all other dimensionless factors,
\begin{equation}
\beta _{p}=\frac{(-1)^{p-1}(D-3)!\,\alpha _{p}}{(D-2p-1)!}\,.  \label{beta}
\end{equation}

The method to prove the formula (\ref{factor}) is a power-expansion of the
function $f$ in the variable $F=A+R=\frac{1}{{\ell _{\mathrm{eff}}^{2}}}%
\,\delta ^{[2]}+R$, which is nothing but the AdS curvature. When
expanded, the auxiliary function becomes
\begin{equation}
f=\sum_{q=1}^{n-1}f_{q}\,F^{q}\,.  \label{f sum}
\end{equation}%
An explicit expression for $f_{q}$ depends on the dimension. Remarkably, in
any dimension, $f_{q}$ is a linear combination of the degeneracy
coefficients $\Delta ^{(p)}$ ,
\begin{equation}
f_{q}=\sum_{p=1}^{q}C_{qp}\,\Delta ^{(p)}\,,  \label{fq sum}
\end{equation}%
where $C_{qp}$ are constant tensors. The last
formula is the key step in finding a generalized conformal mass, because in
general it is not straightforward to prove that only the degeneracy
conditions $\Delta ^{(p)}$ will appear in above expansion.

Eq.(\ref{fq sum}) is always fullfiled in Lovelock theory. In particular, in
the $k$-fold degenerate theories, all $\Delta ^{(p)}$ vanish up to the order $k-1$
by definition, implying that
\begin{equation}
f_{1}=\cdots =f_{k-1}=0\,.
\end{equation}%
In turn, the lower limit in the sum (\ref{fq sum}) changes to
\begin{equation}
f_{q}=\sum_{p=k}^{q}C_{qp}\,\Delta ^{(p)}\,,  \label{key formula}
\end{equation}%
and the first non-vanishing term in the auxiliary function (\ref{f sum})
becomes%
\begin{equation}
f(R)=\sum_{q=k}^{n-1}\sum_{p=k}^{q}C_{qp}\,\Delta
^{(p)}\,F^{q}=C_{kk}\,\Delta ^{(k)}F^{k}+\mathcal{O}(F^{k+1})\,.
\label{Ckk importance}
\end{equation}%
It leads to the polynomial%
\begin{eqnarray}
\mathcal{P}(R) &=&\sum_{q=k}^{n-1}\sum_{p=k}^{q}C_{qp}\,\Delta
^{(p)}\,F^{q-k}  \notag \\
&=&\sum_{s=0}^{n-1-k}\sum_{p=k}^{k+s}C_{k+s,p}\,\Delta
^{(p)}\,F^{s}=C_{kk}\,\Delta ^{(k)}+\mathcal{O}(F)\,.  \label{P final}
\end{eqnarray}

It is important to point out that, in general, the polynomial on the global
AdS space, $F=R+A=0$, is finite, and its contribution in any $D$ is
\begin{equation}
\mathcal{P}(-{A})=C_{kk}\,\Delta ^{(k)}\neq 0\,.
\end{equation}

On the other hand, on the asymptotic boundary, the extrinsic curvature and
the AdS tensor behave as (\ref{extrinsic}) and (\ref{WeylAsymp}),
respectively, so that the charge density tensor at large distances becomes
\begin{eqnarray}
\tau _{j}^{i} &=&\frac{{\theta }_{D}}{16\pi G}\,\delta ^{\lbrack
2n-1]j}\,K_{i}\,F^{k}\mathcal{P}(R)  \notag \\
&=&-\frac{{\theta }_{D}}{16\pi G\,{\ell _{\mathrm{eff}}}}\,\delta
_{i}^{[2n-1]j}F^{k}\mathcal{P}(-{A})+\mathcal{O}(1/r^{D+1})\,,
\end{eqnarray}%
where $\delta ^{\lbrack 2n-1]j}\delta _{i}=\delta _{i}^{[2n-1]j}$. Then,
clearly, because $F^{k}$ behaves as $F^{k}\sim \mathcal{O}(\mu /r^{D-1})$,
it implies that $\tau _{j}^{i}\sim \mathcal{O}(\mu /r^{D-1})$ and therefore $%
Q[\xi ]\sim \mathcal{O}(\mu )$ will give exactly the mass by means of the
charge formula (\ref{Q}). This implies that $\tau _{j}^{i}$ can be
consistently truncated in any dimension by leaving only the leading order of
$F^{k}=W^{k}+\mathcal{O}(1/r^{D+1})$ in the above formula, and all other $%
\mathcal{O}(1/r^{D+1})$ terms can be safely dropped out, such that
\begin{equation}
\tau _{j}^{i}=-\frac{{\theta }_{D}\,C_{kk}\,\Delta ^{(k)}}{16\pi G\,{\ell _{%
\mathrm{eff}}}}\,\delta _{i}^{[2n-1]j}W^{k}\,.  \label{final Tau}
\end{equation}

Remarkably, the result is always proportional to $\Delta ^{(k)}$, and since ${%
\theta }_{D}$ is known, we only have to find the coefficient $C_{kk}$ in a
particular dimension, to be able to carry out the charge evaluation until
the end. In any case, it must be of the form
\begin{equation}
\tau _{i}^{j}\propto \frac{{{\Delta ^{(k)}}}}{16\pi G}\,\delta
_{ii_{2}\cdots {i}_{2k}{i}_{2k+1}}^{jj_{2}\cdots {j}_{2k}{j}_{2k+1}}{W}_{{j}%
_{2}{j}_{3}}^{{i}_{2}{i}_{3}}\cdots {W}_{{j}_{2k}{j}_{2k+1}}^{{i}_{2k}{i}%
_{2k+1}}\,,
\end{equation}%
multiplied by some $D$-dependent real number.

Since the value of $C_{kk}$ strongly depends on the dimension, let us find
now the explicit expressions for $f(R)$ and $\mathcal{P}\left( R\right) $ in
case of the even and odd $D$.

\subsection{Even dimensions}

\label{evenfactorization}

Consider the charge density tensor (\ref{echarge}) in $D=2n$ dimensions of
the form (\ref{tau}), where the scalar coefficient is%
\begin{equation}
{\theta }_{2n}=\frac{{{\ell _{\mathrm{eff}}^{2(n-1)}}}}{{2}^{n-2}}\,,
\label{theta}
\end{equation}%
and the auxiliary function reads
\begin{equation}
f=\sum_{p=1}^{n-1}{\frac{p{\alpha _{p}\ell _{\mathrm{eff}}^{2(1-p)}R^{p-1}}}{%
(2n-2p)!}}\left[ \left( \frac{1}{\ell _{\mathrm{eff}}^{2}}\,\delta ^{\lbrack
2]}\right) ^{n-p}-(-R)^{n-p}\right] .
\end{equation}%
In terms of the variables $A$ and $R$ given by eq.(\ref{a,b}) and redefined
Lovelock coefficients (\ref{beta}), the function can also be written as%
\begin{equation}
f=\frac{1}{2\left( 2n-3\right) !}\sum_{p=1}^{n-1}{\frac{p{\beta _{p}\ell _{%
\mathrm{eff}}^{2(1-p)}}}{n-p}\,}\left[ \left( A-{F}\right) {^{p-1}}%
A^{n-p}-\left( A-{F}\right) {^{n-1}}\right] .
\end{equation}%
The binomials $(A-F){^{p-1}}$ and $(A-F){^{n-1}}$ can be expanded in series,%
\begin{eqnarray}
f &=&\frac{1}{2(2n-3)!}\sum_{p=1}^{n-1}{\frac{p\beta _{p}{\ell _{\mathrm{eff}%
}^{2(1-p)}}}{n-p}}\left( \sum_{q=1}^{p-1}\frac{\left( -1\right) ^{q}\left(
p-1\right) !}{q!\left( p-1-q\right) !}\right.   \notag \\
&&\qquad \qquad -\left. \sum_{q=1}^{n-1}\frac{\left( -1\right) ^{q}\left(
n-1\right) !}{q!\left( n-1-q\right) !}\right) F^{q}A^{n-1-q}\,.
\end{eqnarray}%
In the above expression it was used that, inside the brackets, the terms $q=0
$ cancell out,%
\begin{equation*}
\left( \frac{\left( p-1\right) !}{\left( p-1\right) !}-\frac{\left(
n-1\right) !}{\left( n-1\right) !}\right) A^{n-1}=0\,,
\end{equation*}%
and the sums start from $q=1$.

Obtained power-series in $F$ has the form (\ref{f sum}), but we also have to
show that the corresponding coefficients can be written as a linear
combination of the degeneracy parameters $\Delta ^{(q)}$ defined in eq.(\ref%
{Delta-q}), as given by (\ref{fq sum}).

Indeed, the explicit form of the first few terms of the series is%
\begin{eqnarray}
f &=&\frac{1}{2(2n-3)!}\sum_{p=1}^{n-1}p\beta _{p}{\ell _{\mathrm{eff}%
}^{2(1-p)}}\left( FA^{n-2}-\frac{p+n-3}{2}\,F^{2}A^{n-3}\right.   \notag \\
&&+\left. \frac{p^{2}+np-6p+n^{2}-6n+11}{6}\,F^{3}A^{n-4}+\mathcal{O}%
(F^{4})\right) \,.
\end{eqnarray}%
Using the definition (\ref{Delta-q}),%
\begin{equation}
q!\Delta ^{(q)}\ell _{\mathrm{eff}}^{2(1-q)}=\sum_{p=q}^{n-1}p(p-1)\cdots
(p-q+1)\,\ell _{\mathrm{eff}}^{2(1-p)}\beta _{p}\,,
\end{equation}%
the auxiliar fuction becomes%
\begin{eqnarray*}
f &=&\frac{1}{2(2n-3)!}\left[ \Delta ^{(1)}FA^{n-2}-\left( \frac{n-2}{2}%
\,\Delta ^{(1)}+{\ell _{\mathrm{eff}}^{-2}\,}\Delta ^{(2)}\right)
\,F^{2}A^{n-3}\right.  \\
&&+\left. \left( \frac{n^{2}-5n+5}{6}\,\Delta ^{(1)}+\frac{n-3}{3}\,{\ell _{%
\mathrm{eff}}^{-2}\,}\Delta ^{(2)}+{\ell _{\mathrm{eff}}^{-4}\,}\Delta
^{(3)}\right) \,F^{3}A^{n-4}+\mathcal{O}(F^{4})\right] .
\end{eqnarray*}

In general, for the term $f=C_{qq}\Delta ^{(q)}F^{q}+\cdots $, it can be shown that the
corresponding coefficient is
\begin{equation}
C_{qq}=\frac{\left( -1\right) ^{q-1}\ell _{\mathrm{eff}}^{2(1-q)}{A^{n-1-q}}%
}{2\left( 2n-3\right) !}=\frac{\left( -1\right) ^{q-1}\delta ^{\lbrack 2]}{%
^{n-1-q}}}{2\left( 2n-3\right) !\ell _{\mathrm{eff}}^{2(n-2)}}\,.
\label{cqq}
\end{equation}

Evaluated in the charge density tensor (\ref{final Tau}), it gives%
\begin{equation}
\tau _{j}^{i}=\frac{{{{(-1)}^{k}}\ell _{\mathrm{eff}}\,}\Delta ^{(k)}}{16\pi
G\,2^{n-1}\left( 2n-3\right) !}\,\delta _{i}^{[2n-1]j}\,\delta ^{\lbrack 2]}{%
^{n-1-k}}\,W^{k}\,,
\end{equation}%
or equivalently, with all the indices,%
\begin{eqnarray}
\tau _{i}^{j} &=&\frac{{{{(-1)}^{k}\ell _{\mathrm{eff}}}}\,{{\Delta ^{(k)}}}%
}{16\pi G\,{2}^{n-1}(2n-3)!}\,\delta _{ii_{2}\cdots {i}_{2k}{i}_{2k+1}\dots {%
i}_{2n}i_{2n-1}}^{jj_{2}\cdots {j}_{2k}{j}_{2k+1}\dots {j}_{2n}j_{2n-1}}{W}%
_{{j}_{2}{j}_{3}}^{{i}_{2}{i}_{3}}\cdots {W}_{{j}_{2k}{j}_{2k+1}}^{{i}_{2k}{i%
}_{2k+1}}\times   \notag \\
&&\qquad \qquad \qquad \qquad \qquad \times {\delta }_{{j}_{2k+2}{j}%
_{2k+3}}^{{i}_{2k+2}{i}_{2k+3}}\cdots {\delta }_{{j}_{2n}{j}_{2n-1}}^{{i}%
_{2n}{i}_{2n-1}}\,.
\end{eqnarray}%
By means of the identity%
\begin{equation}
\delta _{ii_{2}\cdots {i}_{2k}{i}_{2k+1}\dots {i}_{2n}i_{2n-1}}^{jj_{2}%
\cdots {j}_{2k}{j}_{2k+1}\dots {j}_{2n}j_{2n-1}}{\delta }_{{j}_{2k+2}{j}%
_{2k+3}}^{{i}_{2k+2}{i}_{2k+3}}\cdots {\delta }_{{j}_{2n}{j}_{2n-1}}^{{i}%
_{2n}{i}_{2n-1}}=2^{n-k-1}\left( 2n-2-2k\right) !\,\delta _{ii_{2}\cdots {i}%
_{2k}{i}_{2k+1}}^{jj_{2}\cdots {j}_{2k}{j}_{2k+1}}\,,
\end{equation}%
we arrive at the result%
\begin{equation}
\tau _{i}^{j}=\frac{{{{(-1)}^{k}{{\ell _{\mathrm{eff}}}}\,{{\Delta ^{(k)}}}%
\left( 2n-2-2k\right) !}}}{16\pi G\,{2}^{k}(2n-3)!}\,\delta _{ii_{2}\cdots {i%
}_{2k}{i}_{2k+1}}^{jj_{2}\cdots {j}_{2k}{j}_{2k+1}}\,{W}_{{j}_{2}{j}_{3}}^{{i%
}_{2}{i}_{3}}\cdots {W}_{{j}_{2k}{j}_{2k+1}}^{{i}_{2k}{i}_{2k+1}}\,.
\label{Teven}
\end{equation}

\subsection{Odd dimensions}

\label{oddfactorization}

In $D=2n+1$ dimensions, the charge density tensor (\ref{qodd}) has the form (\ref{tau}%
), with the scalar factor%
\begin{equation}
\theta _{2n+1}=\frac{\left( -1\right) ^{n}(2n-1)}{2^{3n-4}(n-1)!^{2}}\,,
\end{equation}%
and the auxiliary function%
\begin{equation}
f=\sum_{p=1}^{n}\frac{p\beta _{p}\,\ell _{\mathrm{eff}}^{2(n-p)}}{2n-2p+1}%
\int\limits_{0}^{1}du\,\mathcal{I}_{p}(u)\,.
\end{equation}%
The tensorial quantity $\mathcal{I}_{p}(u)$ is given by eq.(\ref{I}). The
goal is to factorize the function $f$ and expand it as a power series in the
variable $F=R+A$, as given by eq.(\ref{a,b}). With these redefinitions, the
tensorial quantity becomes%
\begin{equation}
\mathcal{I}_{p}=\left( F-A+u^{2}A\right)
^{n-1}-(u^{2}-1)^{n-1}A^{n-p}(A-F)^{p-1}\,.
\end{equation}%
If $F=0$, we have $\mathcal{I}_{p}=0$, as expected. When the binomials $%
\left( F-A+u^{2}A\right) ^{n-1}$ and $\left( A-F\right) ^{p-1}$ are expanded
in series, the integral acquires the form%
\begin{equation}
\mathcal{I}_{p}=\left( \sum_{q=0}^{n-1}\frac{\left( n-1\right) !\left(
u^{2}-1\right) ^{n-1-q}}{q!\left( n-1-q\right) !}-\sum_{q=0}^{p-1}\frac{%
\left( -1\right) ^{q}\left( p-1\right) !(u^{2}-1)^{n-1}}{q!\left(
p-1-q\right) !}\right) A^{n-1-q}F^{q}\,.
\end{equation}%
The $q=0$ terms in the sums cancel out exactly, because
\begin{equation}
\left( \frac{\left( n-1\right) !\left( u^{2}-1\right) ^{n-1}}{\left(
n-1\right) !}-\frac{\left( p-1\right) !(u^{2}-1)^{n-1}}{\left( p-1\right) !}%
\right) A^{n-1}=0\,,
\end{equation}%
and we can drop them out of the sums. Then the integrations in $u$ can be performed
using
\begin{equation}
\int\limits_{0}^{1}du\,\left( u^{2}-1\right) ^{n-1}=\frac{\left( -1\right)
^{n-1}2^{2n-2}(n-1)!^{2}}{(2n-1)!}\,,
\end{equation}%
giving rise to%
\begin{eqnarray}
\int\limits_{0}^{1}du\,\mathcal{I}_{p} &=&\left( -1\right)
^{n-1}2^{2n-2}\left( n-1\right) !\left( \sum_{q=1}^{n-1}\frac{(n-q-1)!}{%
2^{2q}(2n-2q-1)!}\right.   \notag \\
&&-\left. \frac{(n-1)!}{(2n-1)!}\sum_{q=1}^{p-1}\frac{\left( -1\right)
^{q}\left( p-1\right) !}{\left( p-1-q\right) !}\right) \frac{\left(
-1\right) ^{q}}{q!}\,A^{n-1-q}F^{q}\,.
\end{eqnarray}%
Using this result, the auxiliary function becomes%
\begin{eqnarray}
f &=&\left( -1\right) ^{n-1}2^{2n-2}\left( n-1\right) !\sum_{p=1}^{n}\frac{%
p\beta _{p}\,\ell _{\mathrm{eff}}^{2(n-p)}}{2n-2p+1}\,\left( \sum_{q=1}^{n-1}%
\frac{\left( -1\right) ^{q}(n-q-1)!}{2^{2q}q!(2n-2q-1)!}\right.   \notag \\
&&\qquad -\left. \frac{(n-1)!}{(2n-1)!}\sum_{q=1}^{p-1}\frac{\left(
-1\right) ^{q}\left( p-1\right) !}{q!\left( p-1-q\right) !}\right)
A^{n-1-q}F^{q}\,.  \label{fodd}
\end{eqnarray}

Similarly as in the even-dimensional case, the coefficients in the expansion
can be re-arranged so that they are expressed in terms of the characteristic
polynomials $\Delta ^{(p)}$, so that $f$ acquires the form (\ref{Ckk importance}). For
example, the straightforward evaluation gives first few terms as
\begin{eqnarray}
f &=&\frac{\left( -1\right) ^{n}2^{2n-3}\left( n-1\right) !^{2}}{\left(
2n-1\right) !}\sum_{p=1}^{n}p\beta _{p}\,\ell _{\mathrm{eff}}^{2(n-p)}\left(
A^{n-2}F-\frac{2n+2p-5}{4}\,A^{n-3}F^{2}\right.   \notag \\
&&+\left. \frac{4p^{2}+4pn-22p+4n^{2}-20n+33}{24}\,A^{n-4}F^{3}+\mathcal{O}%
(F^{4})\right) \,.
\end{eqnarray}%
Rewriting the coefficient in terms of (\ref{Delta-q}), we find
\begin{eqnarray}
f &=&\frac{\left( -1\right) ^{n}2^{2n-3}\left( n-1\right) !^{2}}{\left(
2n-1\right) !}\left[ \ell _{\mathrm{eff}}^{2(n-1)}\Delta
^{(1)}\,A^{n-2}F\right.   \notag \\
&&-\left( \ell _{\mathrm{eff}}^{2(n-2)}\Delta ^{(2)}+\frac{2n-3}{4}\,\ell _{%
\mathrm{eff}}^{2(n-1)}\Delta ^{(1)}\right) A^{n-3}F^{2}  \notag \\
&&+\left( \ell _{\mathrm{eff}}^{2(n-3)}\Delta ^{(3)}+\frac{2n-5}{6}\,\ell _{%
\mathrm{eff}}^{2(n-2)}\Delta ^{(2)}+\right.   \notag \\
&&\left. +\left. \frac{(2n-3)(2n-5)}{24}\,\ell _{\mathrm{eff}%
}^{2(n-1)}\Delta ^{(1)}\right) A^{n-4}F^{3}+\mathcal{O}(F^{4})\right] .
\end{eqnarray}%
Even though a general $C_{qp}$ in the expansion (\ref{Ckk importance}) is difficult to find,
 we need only the coefficient $C_{qq}$, which can be determined as
\begin{eqnarray}
C_{qq} &=&\frac{\left( -1\right) ^{n-q-1}2^{2n-3}\left( n-1\right) !^{2}}{%
(2n-1)!}\,\ell _{\mathrm{eff}}^{2(n-q)}A^{n-q-1}  \notag \\
&=&\frac{\left( -1\right) ^{n-q-1}2^{2n-3}\left( n-1\right) !^{2}}{(2n-1)!}%
\,\ell _{\mathrm{eff}}^{2}\,\delta ^{[2]n-q-1}\,. \label{cqq-odd}
\end{eqnarray}
Now it is straightforward to evaluate the charge density tensor given by eq.(%
\ref{final Tau}),%
\begin{equation}
\tau _{j}^{i}=\frac{\left( -1\right) ^{k}\ell _{\mathrm{eff}}\,\Delta ^{(k)}%
}{16\pi G\,2^{n-1}(2n-2)!}\,\,\delta _{i}^{[2n-1]j}\delta ^{\lbrack
2]n-k-1}W^{k}\,,
\end{equation}%
what can be written will all indices as
\begin{eqnarray}
\tau _{j}^{i} &=&\frac{\left( -1\right) ^{k}\ell _{\mathrm{eff}}\,\Delta
^{(k)}}{16\pi G\,2^{n-1}(2n-2)!}\,\delta _{ii_{2}\cdots {i}_{2k}{i}%
_{2k+1}\dots {i}_{2n}i_{2n-1}}^{jj_{2}\cdots {j}_{2k}{j}_{2k+1}\dots {j}%
_{2n}j_{2n-1}}\,{W}_{{j}_{2}{j}_{3}}^{{i}_{2}{i}_{3}}\cdots {W}_{{j}_{2k}{j}%
_{2k+1}}^{{i}_{2k}{i}_{2k+1}}\times   \notag \\
&&\qquad \qquad \qquad \qquad \qquad \times {\delta }_{{j}_{2k+2}{j}%
_{2k+3}}^{{i}_{2k+2}{i}_{2k+3}}\cdots {\delta }_{{j}_{2n}{j}_{2n-1}}^{{i}%
_{2n}{i}_{2n-1}}\,.
\end{eqnarray}%
Using the identity%
\begin{equation}
\delta _{ii_{2}\cdots {i}_{2k}{i}_{2k+1}\dots {i}_{2n}i_{2n-1}}^{jj_{2}%
\cdots {j}_{2k}{j}_{2k+1}\dots {j}_{2n}j_{2n-1}}{\delta }_{{j}_{2k+2}{j}%
_{2k+3}}^{{i}_{2k+2}{i}_{2k+3}}\cdots {\delta }_{{j}_{2n}{j}_{2n-1}}^{{i}%
_{2n}{i}_{2n-1}}=2^{n-k-1}\left( 2n-2k-1\right) !\,\delta _{ii_{2}\cdots {i}%
_{2k}{i}_{2k+1}}^{jj_{2}\cdots {j}_{2k}{j}_{2k+1}}\,,
\end{equation}%
the charge density becomes%
\begin{equation}
\tau _{j}^{i}=\frac{\left( -1\right) ^{k}\ell _{\mathrm{eff}}\,\Delta
^{(k)}\left( 2n-2k-1\right) !}{16\pi G\,2^{k}(2n-2)!}\,\delta _{ii_{2}\cdots
{i}_{2k}{i}_{2k+1}}^{jj_{2}\cdots {j}_{2k}{j}_{2k+1}}\,{W}_{{j}_{2}{j}_{3}}^{%
{i}_{2}{i}_{3}}\cdots {W}_{{j}_{2k}{j}_{2k+1}}^{{i}_{2k}{i}_{2k+1}}\,.
\label{Todd}
\end{equation}

Note that, in the formula (\ref{fodd}) for the auxiliary function, the
polynomial in $F$ is only until the order $F^{n-1}$, because the sums in $p$ and $q$
have domains $1\leq q\leq p-1\leq n-1$. This implies that the charge density
contains the coefficients $\Delta ^{(q)}$ only until $\Delta ^{(n-1)}$. However, in the
Chern-Simons theory, all coefficients $\Delta ^{(q)}$ vanish, except $\Delta
^{(n)}\neq 0$, giving
\begin{equation}
\tau _{\mathrm{CS},i}^{j}=0\,.
\end{equation}%
For that reason, the mass of Chern-Simons AdS black hole comes from the term $\tau _{(0)i}^{j}
$, due to presence of the energy gap between the AdS space and the zero mass
black hole, as discussed in Subsection \ref{Odd+CS}.

\bibliographystyle{JHEP}
\bibliography{refmass}

\end{document}